\documentclass[prl,aps,twocolumn]{revtex4}

\usepackage{graphicx}
\usepackage{dcolumn}
\usepackage{bm}

\begin{document}

\preprint{Preview v.4}

\title{Extended Coherence Time with Atom-Number Squeezed Sources}

\author{Wei Li}
\author{Ari K. Tuchman}
\author{Hui-Chun Chien}
\author{Mark A. Kasevich}
\affiliation{Physics Department, Stanford University, Stanford, CA
94305}

\date{\today}

\begin{abstract}

Coherence properties of Bose-Einstein condensates offer the
potential for improved interferometric phase contrast. However,
decoherence effects due to the mean-field interaction shorten the
coherence time, thus limiting potential sensitivity.  In this work,
we demonstrate increased coherence times with number squeezed states
in an optical lattice using the decay of Bloch oscillations to probe
the coherence time.  We extend coherence times by a factor of 2 over
those expected with coherent state BEC interferometry. We observe
quantitative agreement with theory both for the degree of initial
number squeezing as well as for prolonged coherence times.

\end{abstract}

\maketitle

Experimental requirements for precision atom interferometry are well
suited to many of the coherence properties of Bose-Einstein
condensates \cite{Andrews,Brian,Hagley,Gupta,Shin}.  BECs possess
narrower momentum distributions than those of ultra-cold atomic
gases, which removes the need for velocity selection during initial
state preparation. The longer coherence length of a condensate
improves phase contrast, and colder temperatures reduce ensemble
expansion during long interferometer interrogation times.
Furthermore, for confined atom-interferometers
\cite{Prentiss,Kurn,Gunther} which require spatial separation of a
wavepacket in close proximity to a guiding surface, the superfluid
properties of a BEC offer an additional advantage. The mean-field
interaction energy in BECs provides an energy gap to external
excitations, effectively decoupling the atomic proof-mass from the
physical sensor.

On the other hand, the coherence time for BEC interferometry can be
significantly reduced with respect to cold atom sources. Prior to
separation, two linked condensates have relative number fluctuations
which support a well defined relative phase. However, when
separated, the interplay of a large on-site mean-field interaction
with large number variance causes rapid dephasing \cite{You}.  This
concern has been addressed previously by using either dilute
condensates \cite{Sackett} or alternatively, Fermi gases which do
not suffer from density broadening mechanisms \cite{Inguscio2004}.
It is possible, however, to retain some of the benefits of BEC
interferometry while minimizing mean-field induced decoherence. The
generation of atom-number squeezed states from a BEC in an optical
lattice \cite{Orzel, Grei2002} has offered the possibility to create
states with reduced sensitivity to mean-field decay mechanisms.

In this work, we study the characteristic time scale for which an
array of BECs preserves relative phase information after becoming
fragmented, and we observe prolonged coherence times for number
squeezed states. The coherence time is probed through the decay time
of a Bloch oscillation, and we find quantitative agreement with
theoretical predictions.

The theoretical treatment for a BEC in an optical lattice begins
with the Bose-Hubbard Hamiltonian \cite{Jaksch1998b}.

\begin{eqnarray}
H=-\gamma \sum_{i}(\hat{a}_{i+1}^{\dag}\hat{a}_{i
}+H.c.)+\sum_{i}\varepsilon_i\hat{N}_{i}+\frac{g\beta}{2}\sum_{i}\hat{a}_{i}^{\dag}\hat{a}_{i}^{\dag}\hat{a}_{i}\hat{a}_{i}
\label{BH}
\end{eqnarray}
where $g\beta$ represents the mean-field interaction energy and
$\gamma$ is the inter-well tunneling matrix element with both terms
dependent on lattice depth. $\varepsilon_i$ denotes the external
potential term and $\hat{a_i}$ and $\hat{a_i}^\dagger$ represent
single particle annihilation and creation operators respectively at
the $i$th lattice site. $N_i$ denotes local atom number which is
replaced by the central well occupation $N$ in the discussion below.

In the regime of low lattice potential and correspondingly large
tunneling $(N\gamma \gg g\beta)$ the many-body ground state of the
system is described by a superfluid state with local atom number
uncertainty $\sigma{(N)}=\sqrt{N}$.  As the potential barrier is
adiabatically raised, the interplay of the interaction and
tunneling terms renders number fluctuations energetically
unfavorable.  In the Bogoliubov limit, number fluctuations
decrease with increasing lattice potential as
$\sigma_S(N)=(\frac{N^2}{1+N g\beta/2\gamma})^{\frac{1}{4}}$
\cite{Burnett}, with commensurately increasing on-site phase
fluctuations. In the limit of $N\gamma \ll g\beta$, the system
enters the Mott-Insulating regime, where the wavefunction at each
site approximates an atom-number Fock state \cite{Grei2002}.

We fragment this array by frustrating tunneling between adjacent
lattice sites.  In previous work, this has been achieved by
diabatically raising a large potential barrier
\cite{Orzel,Greiner}. In this work, we frustrate tunneling by
applying a large energy gradient across the array.  This method is
preferred since the on-site mean field energy (the third term on
the right of Eq. 1) is unaffected by this process. This sudden
localization is analogous to a beam splitter which instantly
separates an ensemble into two distinct paths. The many-body
ground state does not have time to react to the perturbation, and
therefore an array of independent, localized many-atom states is
formed. The initial array state is no longer in the ground state,
and therefore, the array phase collapses.

Relative phase dispersion between adjacent wells is intuitively
understood by considering an array initially prepared in a shallow
lattice potential.  After fragmentation, each coherent state
$|\alpha\rangle$ can be expanded as a superposition of atom-number
Fock states, $|\alpha\rangle =
\sum_{n}\frac{\alpha^{n}e^{-\frac{1}{2}|\alpha|^{2}}}{\sqrt{n!}}|n\rangle$,
where the phase of each term in the superposition evolves as
$\Delta\phi_{n,i} = \mu(n,\varepsilon_i) t/\hbar$, and $\mu = n
g\beta+\varepsilon_i$ is the local chemical potential.  Each number
term in the superposition evolves at a different rate, leading to
relative phase dispersion with a characteristic decay time $\tau_c$.
The decay of array coherence is seen in the time dependence of the
expectation value of the order parameter $\mid\langle \hat{a}\rangle
\mid=\sqrt{N}e^{-N(g\beta)^2t^2/2}$ where $\tau_c
=(\sqrt{N}g\beta)^{-1}$\cite{You}.  This coherence time can be
prolonged, however, by reducing the number of terms in the initial
atom-number state superposition. For squeezed states with reduced
number uncertainty $\sigma_S(N)$, the coherence time increases as
\begin{equation}
\tau_s = (g\beta\sigma_S(N))^{-1} = (g\beta(\frac{N^2}{1+N
g\beta/\gamma})^{\frac{1}{4}})^{-1} .
\end{equation}

We measure $\tau_c$ for different initial number variances by
studying the decay of coherent Bloch oscillations
\cite{Bloch1928,Ben1996, Arimondo}. An energy gradient
$\varepsilon_i=Ei$ is applied to the array where $E$ is written in
units of energy. This drives an oscillatory response in the
quasimomentum, $q$, of the atomic Bloch state with a period $T =
\hbar/2\pi E$. Although Bloch oscillations are traditionally
observed for conditions where the array is described by bands
delocalized spatially over many lattice sites, they also occur for
spatially localized wavefunctions described in the Wannier-Stark
basis \cite{Subir}.  We isolate the lattice sites when $E \gg
\gamma$ but ensure that $E$ is not too large as to cause particle
loss through Zener tunneling \cite{Brian}. The Bloch oscillation is
observed experimentally by interferometrically following the
evolution of the relative phase between adjacent wells $\Delta \Phi
= q\lambda/2\hbar$, where $\lambda/2$ is the lattice period.
Dispersion in momentum space provides a quantitative indication of
dephasing.

The apparatus used in this experiment has been described in detail
\cite{Orzel}. We load $10^8$ $^{87}$Rb atoms into a time-orbiting
potential (TOP) trap.  Evaporative cooling generates a BEC in the
$|F=2,m_F=2\rangle$ state with $1500$ atoms, density $\rho \sim
10^{12}$ cm$^{-3}$ and temperature $\sim0.2$ $T_c$.  Atom number is
determined with absorptive imaging with $20\%$ shot-to-shot
fluctuations and is consistent with the observed condensate fraction
as a function of temperature.

\begin{figure}[htb!]
\includegraphics[scale = .3]{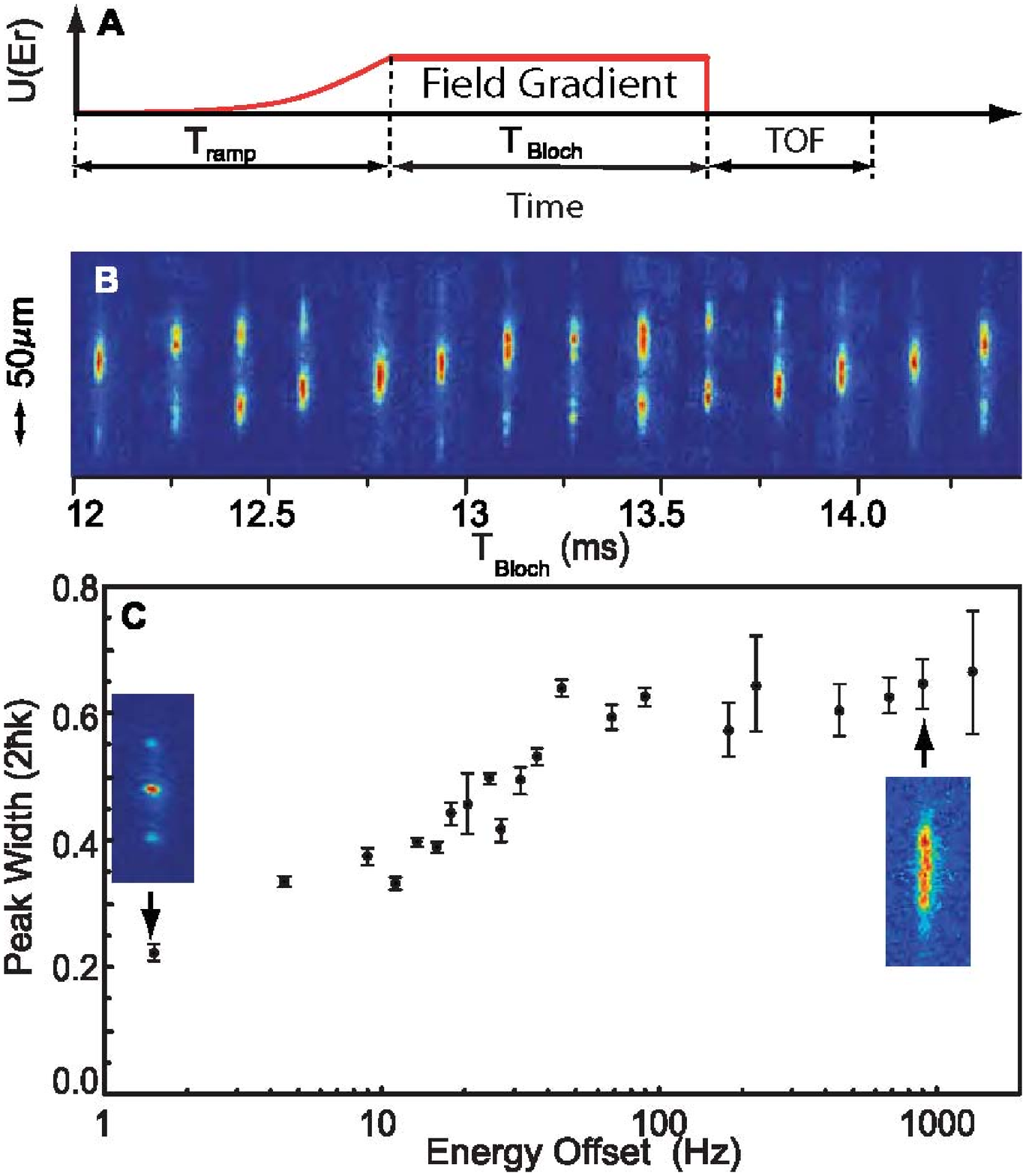} \caption{ \newline {\protect\small
A) Lattice intensity is shown for the experimental sequence. The
lattice intensity is ramped up in T$_{\mathrm{ramp}} = 350$ ms and
then held constant for T$_{\mathrm{Bloch}}$ during which time a
magnetic field gradient is applied.  The lattice and magnetic fields
are turned off and the atoms ballistically expand for a 12 ms Time
of Flight (TOF) before being imaged with a probe pulse.  B)
Absorptive images indicating a Bloch oscillation are shown.  C) Peak
width vs. energy offset $E$ is shown with widths converted to units
of $2\hbar k$. Insets show absorptive images both with and without
dephasing }} \label{Figure1}
\end{figure}

We trap the condensate in a 1-D, vertically oriented optical
lattice. The lattice light is red detuned from the $^{87}$Rb
resonance and has $1/e$ radii of $60$ $\mu$m. The potential depth,
$U$, (measured in $E_R$, where $E_R$ = $\hbar^2 k^2/2m$ and $k =
2\pi/\lambda$ with $\lambda = 852$ nm) is calibrated using three
independent methods which all agree to within $10\%$. We first
measure $U$ by driving an even parity parametric excitation from the
lowest energy band \cite{schori2004}. We also measure the period of
Kapitza-Dirac diffraction by suddenly turning on the optical lattice
\cite{Denschlag2002}.  Finally, we measure the frequency of
Josephson tunneling, which is valid for low lattice depths, $U< 12$
$E_R$ \cite{Inguscio}.  For lattice depths explored in this work
($5<U<24$ $E_R$) we calculate $2\pi\times 4 < \gamma/\hbar <
2\pi\times 250$ Hz, $2\pi\times 0.6< g\beta/\hbar < 2\pi\times 1.8$
Hz and $ 103 < N < 150 $, and the vertical 1/e condensate array
radius ranges between $7-10$ lattice sites.

After state preparation in the optical lattice
(Fig.~\ref{Figure1}A), we drive a Bloch oscillation by applying a
magnetic field gradient along the array axis. We probe the coherence
of the Bloch oscillation by releasing the array, rapidly switching
off the lattice within $500$ ns.  The interferometric signal is
absorptively imaged with single atom detection sensitivity.
Fig.~\ref{Figure1}B shows images depicting a Bloch oscillation with
$T$ = 1.1 ms, taken with $U=10$ $E_R$ and $E/\hbar = 2\pi\times900$
Hz.

We ensure that adjacent sites are decoupled during
T$_{\mathrm{Bloch}}$ by varying $E$ and measuring the width of the
central interference peak.  An increased width indicates dephasing
of the Bloch oscillation. We apply the field gradient for 40 ms
before releasing the array. We see in Fig.~\ref{Figure1}C that for
small $E$ the Bloch oscillation shows minimal dephasing. However,
for large energy offsets the peak width increases, saturating at our
resolution limit for $\gamma\sim E$, where $\gamma/\hbar =
2\pi\times39$ Hz for $U = 12$ $E_R$.

\begin{figure}[htb!]
\includegraphics[scale = .3] {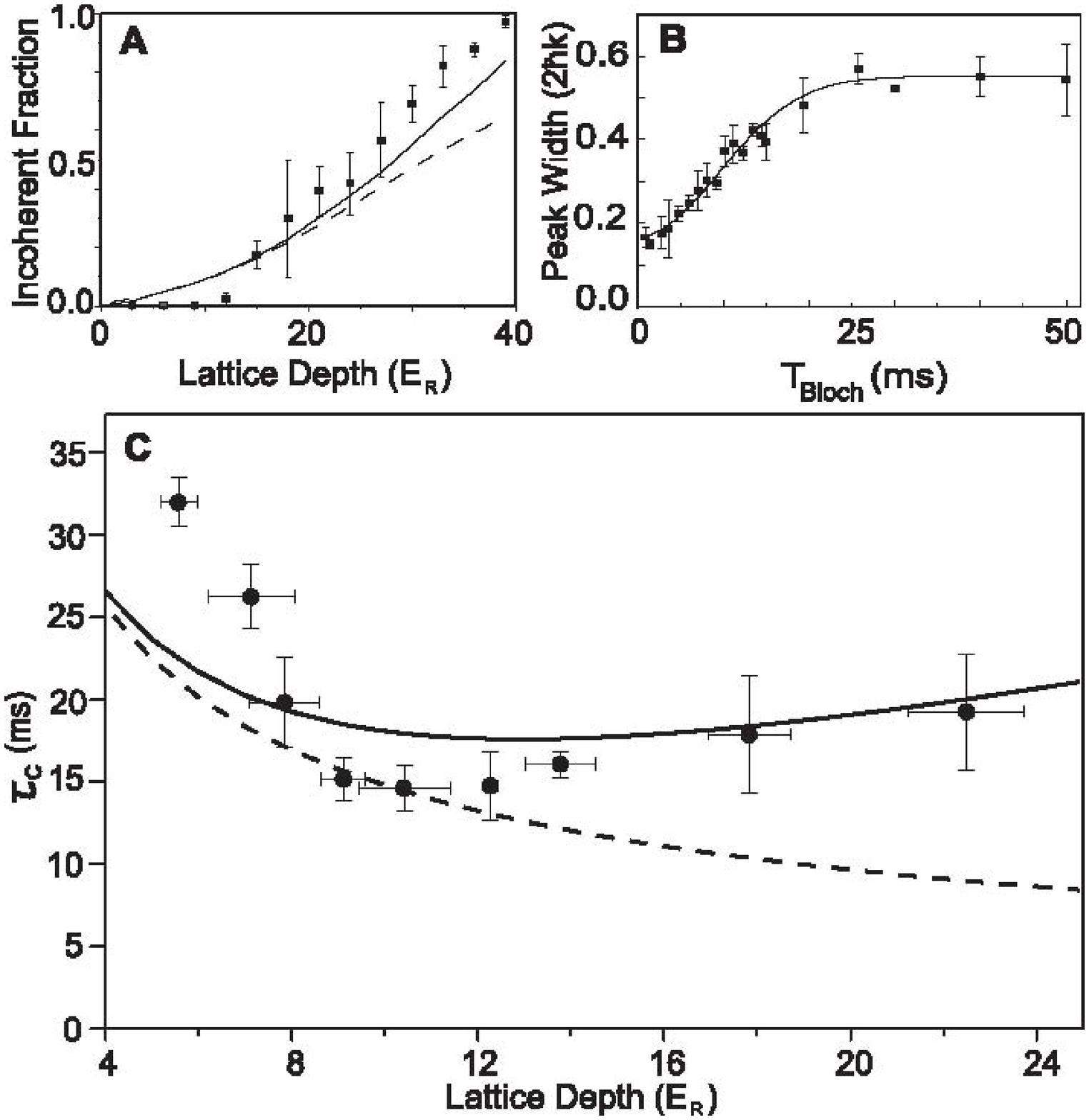}\caption{ \newline {\protect\small
A) The fraction of atoms observed in the incoherent background of
the interference signature is plotted vs. lattice depth. This
fraction from the simulated interference of an array of Gaussian
wavepackets is shown with a solid line.  The calculated quantum
depletion vs. lattice depth is shown with a dashed line.  B) Peak
width is plotted vs. T$_{\mathrm{Bloch}}$ for $U = 10$ $E_R$ and
$E/\hbar = 2\pi\times 900$ Hz. Solid line is a fit to data to
extract $\tau_c = 14.2 \pm 1.3$ ms. C) $\tau_c$ is plotted vs.
lattice depth. Solid (dashed) line denotes theoretical dephasing of
number squeezed (coherent) states. }} \label{Figure2}
\end{figure}

We next quantify the degree of number squeezing generated at a given
lattice depth.  We adiabatically increase the lattice intensity and
then interfere the array, generating an interference pattern with
sharp peaks on top of a broad incoherent background.  We extract
this incoherent fraction by fitting the signal to a function with
three narrow Gaussian peaks and a fourth broad peak. For very low
lattice depths the incoherent background is within our noise floor.
The observed incoherent fraction shown in Fig.~\ref{Figure2}A is a
quantitative metric of number squeezing as seen by comparison with
two theoretical models. First we simulate the interference pattern
from an array of Gaussian BEC wavefunctions with $\sigma_S(N)$
determined by our experimental conditions \cite{Burnett,Orzel}. We
extract the simulated incoherent background (shown with a solid
line) by using the same fit function as used with our experimental
data. Second, we calculate the degree of quantum depletion
\cite{Burnett}, the fraction of atoms not characterized by an order
parameter, expected at a given lattice depth. We equate this
depletion fraction with the incoherent fraction of an interference
signal, shown with a dashed line.  Note that our observed quantum
depletion is significantly higher than that observed in
Ref.~\cite{Xu2006} due to our densities which are nearly two orders
of magnitude lower.

With demonstrated control over number squeezing and site
localization, we explore the dependence of $\tau_c$ on initial
number variance. Using the experimental sequence in
Fig.~\ref{Figure1}A, we take an absorptive image of the interfering
atoms for different lattice depths with $E$ kept constant. We fit
the central vertical peak to a Gaussian to determine the width as a
function of T$_{\mathrm{Bloch}}$. We measure $\tau_c$ by fitting the
data as in Fig.~\ref{Figure2}B to $w(t)=w_f-(w_f-w_0)
e^{-(t/\tau_c)^2}$ \cite{You}. $w_f$ is the maximum observed width
representing a fully dephased signature, and $w_0$ is the peak width
prior to any phase dispersion.

Fig.~\ref{Figure2}C shows the summary graph plotting $\tau_c$ vs.
lattice depth, where increased lattice depth reflects increased
number squeezing. The theoretical $\tau_c$ for isolated coherent
states and number squeezed states (Eq. 2) are shown with dashed and
solid lines respectively.  For very low lattice depths $\tau_c$ is
longer then expected for either coherent state or squeezed state
dephasing. This is likely due to insufficient isolation between
wells when $\gamma$ is large. However, we cannot significantly
increase $E$ at this lattice depth without introducing Zener
tunneling losses. For intermediate lattice depths with
near-Poissonian number variance, we observe good correlation of
$\tau_c$ with that expected for isolated coherent state condensates.

For deeper lattice depths, however, we measure long coherence times
which are in quantitative agreement with theoretical number squeezed
state dephasing. For number squeezed states prepared at $U = 22.5$
$E_R$, $\tau_c = 19.3 \pm 3.5$ ms.  This represents an increase of a
factor of 2.1 over the expected decay time of an array of coherent
states in the same lattice potential.  It is interesting to note
that here squeezing extends the coherence time; typically, the
enhanced fragility of squeezed states to loss mechanisms results in
reduced coherence times \cite{Stockton}.

\begin{figure}[htb!]
\includegraphics[scale = .3]{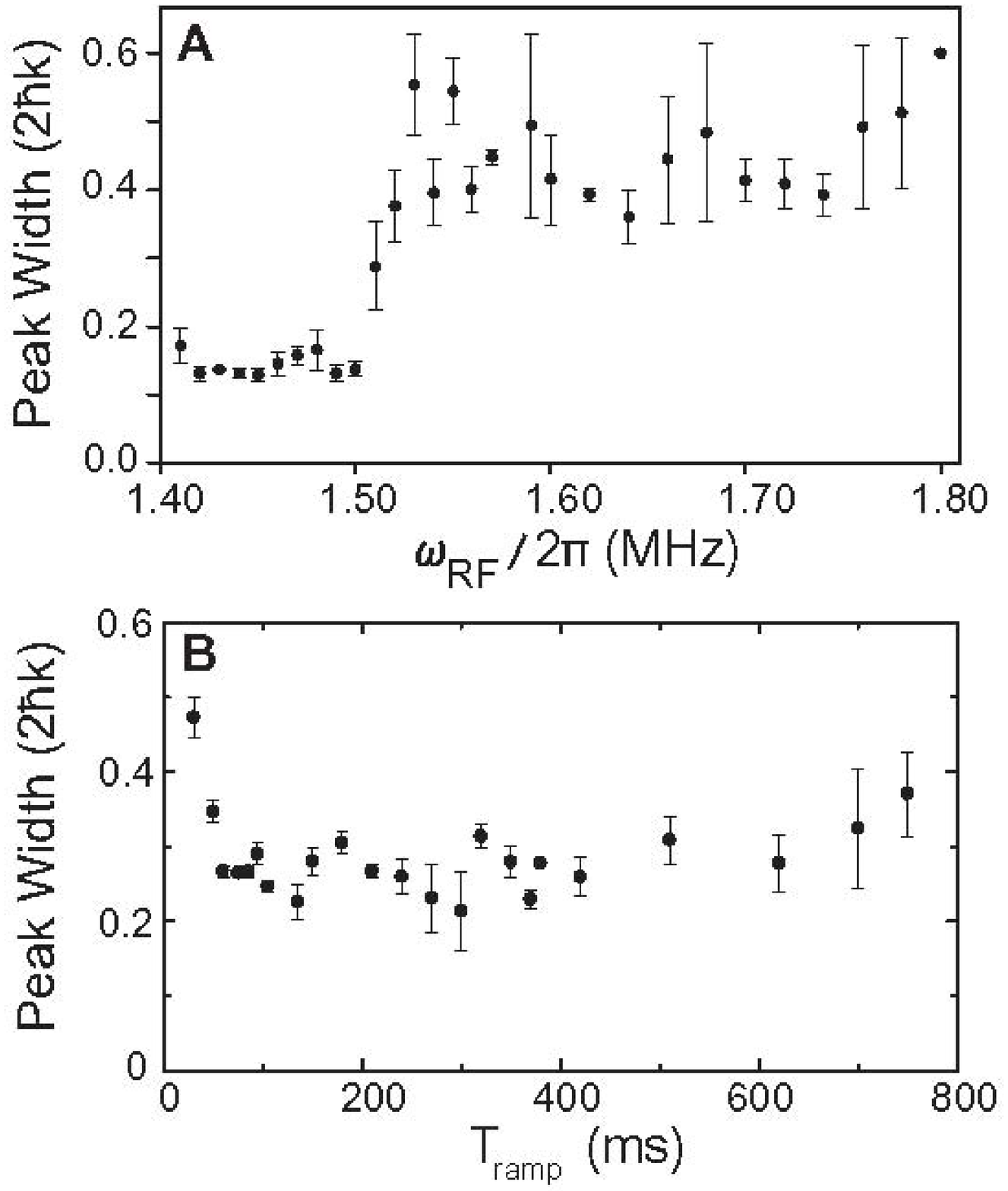} \caption{ \newline {\protect\small
A) Peak width vs. $\omega_{\mathrm{RF}}/2\pi$.  $T_c$ = 1.6 MHz for
a BEC in the bare harmonic trap.  B) Peak width vs.
T$_{\mathrm{ramp}}$
 }} \label{Figure3}
\end{figure}

To eliminate other potential sources of dephasing, we investigate
the effects of finite temperature and the adiabaticity of our
lattice ramp on the interferometric peak width observed in Bloch
oscillations.  We prepare condensates at different temperatures by
varying the final frequency $\omega_{\mathrm{RF}}$ of the
evaporative cooling stage. As seen in Fig.~\ref{Figure3}A, we
observe no change in peak width for
$2\pi\times1.40<\omega_{\mathrm{RF}}<2\pi\times1.50$ MHz with $U=16$
$E_R$. However, at $\omega_{\mathrm{RF}}=2\pi\times1.51$ MHz we
observe a sudden onset of phase broadening.  To avoid this thermal
dephasing regime all data was taken with
$\omega_{\mathrm{RF}}=2\pi\times1.44$ MHz. Note that this critical
temperature in the lattice is different from the BEC transition
temperature in a bare harmonic trap (corresponding to
$\omega_{\mathrm{RF}}=2\pi\times1.6$ MHz).

In Fig.~\ref{Figure3}B, we investigate the dependence of peak width
on lattice intensity ramp speed.  A balance is required to avoid
losses due to lattice heating with very long ramp times and
non-adiabaticity effects with short ramp times. We find that peak
width is insensitive to ramp speed for $80 <$ T$_{\mathrm{ramp}}
<620$ ms.

\begin{figure}[htb!]
\includegraphics[scale = .3]{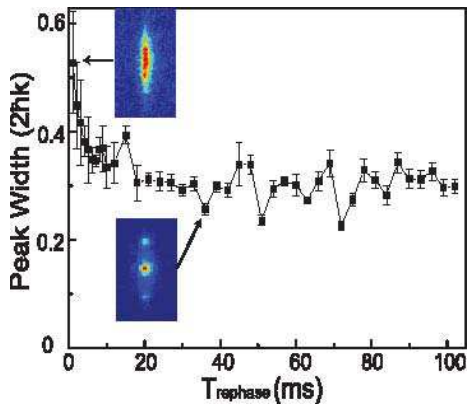} \caption{ \newline {\protect\small
Peak width is plotted vs. continued hold time in lattice after
removing the external field gradient, T$_{\mathrm{rephase}}$.
Insets are absorptive images showing initial dephasing of the phase
contrast and subsequent revival.
 }}
\label{Figure4}
\end{figure}

As a final consideration, we explore the possibility for coherence
restoration after complete dephasing.  We use the same experimental
sequence as in Fig.~\ref{Figure1}A, with $U = 10$ $E_R$ and
T$_{\mathrm{Bloch}}$ = $80$ ms, ensuring that dephasing has
occurred. This time, however, after turning off the magnetic field
gradient we continue to hold the atoms in the optical potential
before interfering them.  We see phase contrast return nearly
completely after a rephasing time $\tau_r \sim 10$ ms as shown in
Fig.~\ref{Figure4}. While the details of this rephasing mechanism
require further investigation, a two-well model predicts this time
to be determined by the generalized Josephson frequency $\omega_J =
\sqrt{N g \beta \gamma}$.  Our observed rephasing time is in
excellent agreement with this prediction for our parameters
\cite{Smerzi}.

In conclusion, we have demonstrated that number squeezed states in
an optical lattice can extend coherence times for interferometry. We
have obtained quantitative agreement with theory both in calibrating
the number variance of our initial state as well as in observed
coherence times with number squeezed states. For sensitivity
considerations in practical interferometers, however, note that
extended coherence times come with the price of increased phase
noise due to number squeezing.  Future work will explore this
tradeoff in optimizing absolute sensor performance.

We acknowledge funding support from DARPA and the MURI on Quantum
Metrology sponsored by ONR.  The work of A.K.T. was supported by an
IC fellowship, sponsored by NGA.


\begin{thebibliography} {}%
\bibitem{Andrews} M. Andrews, {\it et al.}, Science {\bf275}, 637 (1997).

\bibitem{Brian} B. Anderson and M. Kasevich, Science {\bf282}, 39 (1998).

\bibitem{Hagley} Y. Torii, {\it et al.}, Phys. Rev. A {\bf 61}, 041602
(2000).

\bibitem{Gupta} S. Gupta, K. Dieckmann, Z. Hadzibabic, D. Pritchard, Phys. Rev. Lett. {\bf 89},
140401 (2002).

\bibitem{Shin} Y. Shin, {\it et al.}, Phys. Rev.
Lett. {\bf 92}, 050405 (2004).

\bibitem{Prentiss} Y. Wang, {\it et al.}, Phys. Rev.
Lett. {\bf 94}, 090405 (2005).

\bibitem{Kurn} S. Gupta, K. Murch, K. Moore, T. Purdy, D.
Stamper-Kurn, Phys. Rev. Lett. {\bf 95}, 143201 (2005).

\bibitem{Gunther} A. G\"{u}nther, {\it et al.}, Phys. Rev.
Lett. {\bf 95}, 170405 (2005).


\bibitem{You} A. Imamoglu, M. Lewenstein and L. You, Phys. Rev. Lett. {\bf78}, 2511 (1997).

\bibitem{Sackett} J. M. Reeves, {\it et al.}, Phys. Rev.
A {\bf 72}, 051605R (2005).

\bibitem{Inguscio2004} G. Roati, {\it et al.}, Phys. Rev.
Lett. {\bf 92}, 230402 (2004).

\bibitem{Orzel} C. Orzel, A. K. Tuchman, M. L. Fenselau, M. Yasuda and M. A. Kasevich,
Science {\bf291}, 2386 (2001).

\bibitem{Grei2002} M. Greiner, O. Mandel, T. Esslinger, T. Hansch and I.
Bloch, Nature {\bf415}, 39 (2002).

\bibitem{Jaksch1998b} D. Jaksch, C. Bruder, J. I. Cirac, C. W. Gardiner and P. Zoller,
Phys. Rev. Lett. {\bf 81}, 3108 (1998).

\bibitem{Burnett}
K. Burnett, M. Edwards, C. Clark and M. Shotter, J. Phys. B {\bf35},
1671 (2002).

\bibitem{Greiner} M. Greiner, O. Mandel, T. Esslinger, T. Hansch and I.
Bloch, Nature {\bf419}, 51 (2002).

\bibitem{Bloch1928} F. Bloch, Z. Phys. {\bf52}, 555 (1928).

\bibitem{Ben1996} M. Ben Dahan, E. Peik, J. Reichel, Y. Castin and C. Salomon,
Phys. Rev. Lett. {\bf76}, 4508 (1996).

\bibitem{Arimondo} O. Morsch, {\it et al.}, Phys. Rev.
Lett. {\bf 87}, 140402 (2001).

\bibitem{Subir} S. Sachdev, K. Sengupta, S. Girvin, Phys. Rev.
B {\bf 66}, 075128 (2002).

\bibitem{schori2004}
C. Schori, T. Stoferle, H. Moritz, M. Kohl, and T. Esslinger, Phys.
Rev. Lett. {\bf 93}, 240402 (2004).

\bibitem{Denschlag2002}
J. Denschlag, {\it et al.}, J. Phys. B {\bf35}, 3095 (2002).

\bibitem{Inguscio} F. S. Cataliotti, {\it et al.},
Science {\bf 293}, 843 (2001).

\bibitem{Xu2006}
K. Xu, {\it et al.}, Phys. Rev. Lett. {\bf96}, 180405 (2006).

\bibitem{Stockton} J. Stockton, J.M. Geremia, A. Doherty, H.
Mabuchi, Phys. Rev. A {\bf67}, 022112 (2003).

\bibitem{Smerzi} A. Smerzi, S. Fantoni, S. Giovanazzi, S.R. Shenoy, Phys. Rev.
Lett. \textbf{79}, 4950 (1997).


\end{thebibliography}
\end{document}